\documentstyle[12pt,aasms]{article}

\def\la{\mathrel{\mathpalette\fun <}}

\def\fun#1#2{\lower0.837ex\vbox{\baselineskip0ex\lineskip0.209ex
  \ialign{$\mathsurround=0ex#1\hfil##\hfil$\crcr#2\crcr\sim\crcr}}}

\def\msun{M_\odot}

\def\sles{\lower2pt\hbox{$\buildrel {\scriptstyle <}
   \over {\scriptstyle\sim}$}}
 
\def\sgreat{\lower2pt\hbox{$\buildrel {\scriptstyle >}
   \over {\scriptstyle\sim}$}}

\def\la{\mathrel{\mathpalette\fun <}}

\begin{document}

 \title{ The $V-EUV$ Delay for Dwarf Nova Outbursts:
   A Case Study for VW Hydri, U Geminorum, and SS Cygni }

 \vskip 1truein

 \author{ John K. Cannizzo    }
 \affil{e-mail: cannizzo@stars.gsfc.nasa.gov}
 \affil{NASA/GSFC/Laboratory for High Energy Astrophysics, 
 Emergent Information Technologies, Inc.,
 Code 662, Greenbelt, MD 20771}
 \authoraddr{NASA/GSFC/Laboratory for High Energy Astrophysics, 
 Emergent Information Technologies, Inc.,
 Code 662, Greenbelt, MD 20771}

 \vskip 3truein

\vskip 0.5truein
\centerline{ to appear in the Astrophysical Journal,
  2001, August 1, vol. 556 }
\received { 2000 November 16 }
\revised {  2001 February 19  }
\accepted{  2001 March 22 }

\begin{abstract}

We present a parameter study using
time dependent calculations of the thermal
limit cycle model for dwarf nova outbursts.
Our goal is to delineate the dependence
of the delay between
  the initial rapid rise of the
     visual and $EUV$ fluxes
during the start of an outburst
on model parameters,
  concentrating on three
bright, nearby systems for which complete
optical and $EUV$ observations
exist $-$
VW Hyi, U Gem, and SS Cyg.
For each system we compute 15 models
spanning the early part of an outburst,
taking the ratio of the instigation
radius to the outer disk radius to be
either 0.5, 0.7, or 1.0,
and adopting a value for the alpha
viscosity parameter in the ionized
disk of 0.1, 0.15, 0.2, 0.25, or 0.3.
  We confirm Smak's  findings
which show a consistency of the standard
model with observations.
For these systems we infer that the outburst
must  be triggered at $\sim0.7-0.8$ of the 
outer disk radius to produce delays which are
in accord with observations.
We show that the level of the $EUV$ 
flux in outburst is dictated by the
amount of material carried within the inward
moving heating front spike as it reaches the inner
edge of the disk, and we re-affirm earlier work
by Meyer and Lin et al. which found that the heating
front speed is given by the alpha parameter times
the sound speed within the heating spike.
 We also see a stagnation or period of slow
warming (first noted by Mineshige)
   during the early stages of thermal instability,
but find it does not influence the $V-EUV$ delay
since it precedes the rapid rise in     $V$ 
at the start of the outburst.
   In studying the sensitivity of our results
to the initial mass distribution,
we find that if one decreases the surface density
in the inner disk, interior to the instigation
radius,
the $V-EUV$ delay
can be lengthened by as much as a factor of two.
  In addition,
     we find there to be a weak
relation between the $V-EUV$ delay  and the value
of the alpha viscosity   parameter in quiescence.

\end{abstract}


\section{ INTRODUCTION }

The accretion disk  limit cycle model
was proposed        twenty years ago to
account for the semi-periodic outbursts
seen in dwarf novae (Meyer \& Meyer-Hofmeister 1981).
Dwarf novae are a subclass of the cataclysmic variables
   (CVs)
    $-$ interacting binary stars in which a Roche
lobe filling K or M star transfers material through
the inner Lagrangian point into an accretion disk 
about a white dwarf (WD) primary (Warner 1987, 1995).
 Dwarf novae are defined by their outbursts, which recur
on intervals of between days and decades, and have 
amplitudes of several magnitudes.

The limit cycle model posits that the rate of
mass transfer into the outer part of the disk
from the mass losing star  is constant on   long time
scales.
(Actually in the AM
Her subclass of CVs, for which the WD is so 
strongly magnetized that no disk can form,
   one sees strong, chaotic variations
  in the mass transfer rate from the
  secondary star
   [see Fig. 3 of Schreiber,
 G\"ansicke, \& Hessman 2000].
 The mass losing
  secondary 
    stars in CVs
  should not differ between subclasses,
so highly  variable mass transfer
should also be the rule in dwarf novae. 
For the non-magnetic systems however,
the accretion disk about the WD
     acts like a low-pass filter which
responds only to slow variations $-$
those occurring  on time scales  that  are long
compared to the viscous time in the outer disk  $\sim 1$ yr.)
   The accretion disk limit cycle model is based
on the theoretical  finding that the steady state relation between
effective temperature $T_{\rm eff}$ and surface density $\Sigma$
at a given radius forms a hysteresis with two stable branches $-$
one for ionized  gas and one for neutral gas.  When a series
of steady state solutions for a given radius and varying
rate of accretion ${\dot M}$ are plotted as $T_{\rm eff}$ versus
$\Sigma$, one sees an S-shaped curve.
     The basic idea is that in the quiescent state of dwarf
novae, matter is accumulating and the mass of the accretion
disk is increasing. The mass flow within the disk is highly
non-steady, and the rate of removal of mass from the inner
disk is many orders of magnitude below that at which it is being
fed at the outer edge. In the outburst state the  opposite holds $-$
the rate of mass removal from the inner disk onto the central
accretor exceeds the rate of mass  supply to the outer disk edge
which results in a decrease of disk mass.  Thus, 
   both the high and   low state disks are    unstable 
   in the sense that they 
     continually attempt  to
revert to the opposite state.
For recent  reviews
  see
      Osaki (1996),   Cannizzo (1993a, 1998c).
%
%
%
Following the initial vertical structure papers
 which established
the physical underpinnings for the limit cycle idea
(Meyer \& Meyer-Hofmeister 1981,
 Meyer \& Meyer-Hofmeister 1982,
Cannizzo, Ghosh, \& Wheeler 1982,
Cannizzo \& Wheeler 1984,
Smak 1984,
Faulkner, Lin, \& Papaloizou 1983,
Mineshige \& Osaki 1983),
workers began to explore detailed aspects of the model
using time dependent codes
  (Papaloizou, Faulkner, \& Lin 1983, Lin, Faulkner, \& Papaloizou
  1985, Smak 1984, Meyer \& Meyer-Hofmeister 1984, Mineshige \& Osaki
1985, Cannizzo, Wheeler, \& Polidan 1986=CWP, Cannizzo \& Kenyon 1987=CK,
Pringle, Verbunt, \& Wade 1986=PVW).
       More recent time dependent
 work has concentrated on the development
of increasingly sophisticated
 numerical models which are utilized
to study different systematic
effects inherent in the calculations
(e.g., Mineshige 1988, Ichikawa \& Osaki 1992, 1994,
 Cannizzo 1993b=C93b, 
Ludwig, Meyer-Hofmeister, \& Ritter 1994,
Ludwig \& Meyer 1998,
Hameury et al. 1998,
Menou,  Hameury, \& Stehle 1999,
Truss et al. 2000,
Buat-Me\'nard,  Hameury, \& Lasota 2001=BHL).
          These later studies represent
an advancement over the  earlier
ones,  but
some of the difficulties
in reconciling theory with observation remain.
For instance,
 BHL present global time dependent
computations
(using the model described  in Hameury et al. 1998)
which include the effects
 of heat input due to the stream-disk
impact
and tidal  torque.
They show that such models produce
sequences of 
alternating 
 long and short outbursts triggered at large
radii that have similar peak luminosities,
thereby rectifying
some of the difficulties
evident in the  computations of C93b.
Figure 7 of BHL shows a change in sequencing
from all short to all long outbursts
as the  secondary   mass transfer rate ${\dot M}_T$
is made to increase,
while their Figure 4
shows a decrease in the inter-outburst
quiescent intervals with increasing ${\dot M}_T$.
Combining these results  gives a prediction
that $<N(L)/N(S)>$  $ - $
the moving long term average of the ratio
of number of long to short outbursts
in a   long time series (during which 
there are variations 
      in   ${\dot M}_T$)    $-$
should  be anti-correlated with $<t_q>$  $-$
the moving long term average
of the quiescent interval.
In other words,
   if long term variations
in  $<{\dot M}_T>$
occur and affect the outbursts, then
during times when
 $<{\dot M}_T>$ is larger
and  therefore  more long outbursts are occurring,
the quiescent intervals
should be shorter.
The long term light curve
of SS Cygni
shows the opposite:
a strong positive
correlation between
   $<N(L)/N(S)>$ and $<t_q>$
(see Figs. 14 and 15 of C93b).

  One observation of interest is the delay between the rise
of the optical and $EUV$ fluxes in so-called ``fast-rise''
outbursts which are seen in several dwarf novae.
For these systems one sees the optical begin to rise about
1 d before the $EUV$. 
This is accounted for
in the model 
 by the
fact that the optical flux, which is weighted basically by the
emitting area of the disk, arises  predominantly at large radii,
while 
     the $EUV$ flux comes from 
     small radii close to the WD
  (Cannizzo 1996, 1998a).
       Therefore the
delay represents the time for the heating transition wave
  which communicates the onset of thermal instability
to travel from large radii where the instability begins down to
 the WD.
  The earliest studies to explore the $V-EUV$ delay found a
basic  consistency  between theory and observation (Smak 1984,
  CWP), while later workers
found an apparent discrepancy (PVW, CK).
 Mineshige (1988) presents  a detailed time dependent
study of the rise to outburst
 using more sophisticated input
   physics than some of the earlier workers.
  In particular, he uses  physically realistic values of
the specific heat
      as a function of density and temperature  $c_P(\rho,T)$,
  and finds a ``stagnation'' stage in  the early phase of
the outburst during which the disk  midplane temperature
is stalled at $\sim10^4$ K due to the large $c_P$.
  Some other aspects of his model rely   on
detailed $ T_{\rm eff}(\Sigma)$ features
  near the local maximum in surface density
            $\Sigma_{\rm max}$
     which depend strongly  on
specifics of the vertical structure calculations.

    Smak  (1998) presents  a
thorough investigation of the problem, and traces the
failure of the later workers to their inability to produce
outbursts which begin at sufficiently large radii.
    The reason this is critical is 
at least partly due to the fact
that outbursts which 
begin too close to the inner edge, referred to as ``inside-out'' outbursts
  (CWP) or ``type B'' outbursts  (Smak 1984),
    have a fundamentally
different shape  than outbursts which begin at large radii,
 which are 
     referred to as  ``outside-out'' outbursts
   (CWP) or ``type A'' outbursts (Smak 1984).
 Basically, type A outbursts have a fast rise, and type B
outbursts have a slow rise. The difference 
     has to do with the rate
of enhancement of surface density in the inner disk
during the early stage of the outbursts (see CWP for an
explanation). Furthermore, the dividing line between these
two types of outbursts is quite sharp $-$
for a triggering
  or instigation
    radius $r_{\rm instig}$ less than
some critical radius  $r_{\rm crit}$
      one obtains type B outbursts,
 whereas for $r_{\rm instig} > r_{\rm crit}$, type A outbursts occur
(see Fig. 6b of Cannizzo 1998b).
In summary, 
  in order to obtain the needed
  delay  one requires
 an outburst which triggers 
     at  a sufficiently large radius
            so that
  it is   not only type A,
  but also that
the time for the heating front to travel
to the inner disk edge
$\int dr |v_F|^{-1}$ 
     is long enough to equal the observed
$\Delta t(V-EUV)$.

      Smak (1998) shows that the flux distributions
  utilized in the calculations of the $V$ and $EUV$
fluxes
(i.e. Planckian versus stellar or Kurucz-type)
    are of secondary importance, in contrast to what
might be inferred from some reviews (e.g. 
     Cannizzo 1993a, Cannizzo 1998c).
This is particularly true in comparing the $V$ and $EUV$
fluxes: in fact for the latter one may simply use
${\dot M}_{\rm inner}$  $-$
the rate of mass loss from the inner disk onto the WD $-$
as a tracer of the $EUV$ flux.       
In this regard the  {\it Extreme Ultraviolet Explorer}
satellite ({\it EUVE})  is an ideal instrument
  for studying accretion onto WDs:
 the peak in its response function lies at $\lambda\simeq100$ A
or $E\simeq 0.1$ keV, 
and the effective temperature in the disk
$T_{\rm eff}
= T_{*}
 (r_{\rm WD}/r)^{-3/4}
    (1-\sqrt{r_{\rm WD}/r})^{1/4}$, 
where $\sigma {T_{*}}^4  =
   (3/8\pi)(GM_{\rm WD}/r_{\rm WD}^3) {\dot M} $
(Shakura \& Sunyaev 1973).
Evaluating gives   $T_{*} 
 =  18.7 $ eV
$({\dot M}/10^{18}$ g s$^{-1})^{1/4}$
$(M_{\rm WD}/1\msun)^{1/4}$
$(r_{\rm WD}/5\times 10^8$ cm$)^{-3/4}$. 
 The maximum temperature 
 in  the accretion
disk     (evaluated
at $[49/36]r_{\rm WD}$) is $0.488T_{*} =  9.1$  eV
$({\dot M}/10^{18}$ g s$^{-1})^{1/4}$
$(M_{\rm WD}/1\msun)^{1/4}
  (r_{\rm WD}/5\times 10^8$ cm$)^{-3/4}$.
 Therefore the peak in the accretion
  disk spectrum
occurs at about a factor of ten lower  in energy
than the {\it EUVE} response peak.
This region of the spectrum is largely masked
by interstellar absorption.
  The flux seen by {\it EUVE}
for dwarf novae in outburst must be
predominantly coming from the boundary layer
around the WD.   
  We idealize this flux as representing
the mass flux directly onto the WD from the
inner edge of the accretion disk.

   The
    perception of a ``UV delay problem''
led many workers to consider scenarios for the quiescent
evolution of dwarf novae in which the surface density
is eroded at small radii via some type of low-level
accretion onto the WD such that one creates a hole
in the inner disk, thereby   forcing the outbursts to
be triggered at larger radii than would have otherwise
happened (e.g. Livio \& Pringle 1992,
    King 1997, Stehle \& King 1999,
  Hameury, Lasota, \& Dubus 1999).  
  Although this specific 
     impetus for such models is rendered 
     invalid
by Smak's study, there may  be other reasons
  to invoke some quiescent evaporation of the disk
  in dwarf novae, such
as the observations of significant X-ray flux coming from
the central star in quiescence, which cannot be explained
in the standard model (e.g. Mukai et al. 1997,
van Teeseling 1997, Pratt et al. 1999).

In this work we expand on Smak's (1998) investigation
 by carrying out a detailed parameter study
   for three specific systems
using a time dependent model for the development
of the onset and subsequent evolution of the accretion
disk following the start of a dwarf nova outburst.
  We 
use parameters appropriate for VW Hyi, U Gem, and SS Cyg
because 
for these
dwarf novae there exist simultaneous $V$
and $EUV$ data
 during outburst 
(Mauche, Mattei, \& Bateson 2001=MMB).
  The  $V$ data in MMB
      were obtained by the
American Association of Variable Star Observers
(AAVSO) and the
 Variable Star Section/Royal Astronomical
 Society of New Zealand (VSS/RASNZ);  
       their $EUV$ data comes from {\it EUVE}.
We perform a parameter study over $\alpha_{\rm hot}$
$-$
the viscosity parameter characterizing the hot state of
the disk, and $r_{\rm instig}/r_{\rm disk}$ $-$
the ratio  of the triggering radius for thermal
instability to the outer disk radius.
   We also examine the sensitivity of the
results
 to the initial mass profile in the disk.
Specifically,  we examine (i)
       the effect of putting
all the initial mass into a torus at
 $r_{\rm instig}$ and leaving only 
       a small surrounding
background surface density $\Sigma_{\rm floor}$,
      and
 (ii) the effect of  varying  $\alpha_{\rm cold}$
which governs the quiescent disk mass.



\section { MODEL CALCULATIONS }

\subsection { Numerical Model   }

We utilize a new version of our standard
time dependent disk instability code
which was recently rewritten from FORTRAN
into C++. The code has been tested to ensure
numerical stability.
  The time dependent equations we solve
are given in C93b.
  One combines the mass and angular
momentum conservation
equations into a diffusion equation
    for surface density,
  written under the assumptions of Keplerian
rotation $v_{\phi}(r)=(GM_{\rm WD} r^{-1})^{1/2}$
     and cylindrical symmetry.
     One similarly writes an energy equation
 under the assumption of cylindrical symmetry.
Together these      two coupled nonlinear
differential equations 
        provide a means for calculating small
changes in the surface density $\delta\Sigma(r,t)$
and midplane temperature $\delta T(r,t)$
 for a given physical state of the disk
  specified by $\Sigma(r,t)$ and $T(r,t)$.

   If  the disk
   is entirely on either the lower
or upper stable branch of solutions
(i.e., completely neutral gas or completely ionized gas),
      then only the
diffusion equation for $\Sigma(r,t)$ is solved
 in that time step.
   For such solutions we utilize power law
scalings for $T(\Sigma,r,\alpha)$ 
  given in C93b 
   to determine the
midplane temperature.
   For radii in which thermal transitions are
occurring, we must also solve the energy equation,
  which contains both direct heating and cooling
  terms characterizing the departure from thermal
equilibrium in a given annulus, as well as terms
 involving advection (the physical transport
 of heat by the flow) and terms representing
 the radial energy flux
within the body of the disk.
 As in C93b,
     we interpolate $\alpha$
logarithmically 
between $\alpha_{\rm cold}$
and $\alpha_{\rm hot}$
 based on the local value of $T$
 compared to $T(\Sigma_{\rm min})$ and  
             $T(\Sigma_{\rm max})$.
 The advective terms contain the local
flow velocity
        $v_r = -(3\nu/r)$
$\partial\log(\nu\Sigma r^{1/2})/\partial\log r$
which is calculated through (upstream)
   differencing at each grid point.
 Two terms contribute to the
radial flux of energy,
a viscous term 
$J_V=(3/2)(c_P\nu\Sigma/r)$
$ \partial(r\partial T/\partial r)/\partial r $
  (introduced by Mineshige 1986)
    and a radiative term
 $J_R = -(h/r) \  \partial(rF_r)/\partial r $,
where $F_r = -[4acT^3/(3\kappa\rho)] \partial T/\partial r $
                     (introduced by Faulkner et al. 1983).
In C93b the radiative term $J_R$
was not utilized due to problems with numerical stability.
In this work we have included this term, using the
analytical scalings for the opacity $\kappa$  given
in    Lin \& Papaloizou (1985).
      As we shall see
     later, $|J_R/J_V|<<1$ in general,
 which is ironic given that
  the introduction of  $J_R$ preceded
that of $J_V$.

 We take $N=1000$
grid points equally spaced in $\sqrt{r}$
between $r_{\rm WD }$ and  $r_{\rm disk}$.
 For the inner disk radius $r_{\rm WD }$
       we use the 
analytical white dwarf radius-mass scaling
due to B. Paczy\'nski, as given by Anderson (1988,
  see his eqn. [33]).
We specify an initial surface density profile 
      so as to force the triggering for outburst
at a given radius.
This is  accomplished by starting with a Gaussian profile
for $\Sigma(r)$ centered at some $r_c$
which is either 0.5, 0.7, or 1.0 of the outer disk radius 
  $r_{\rm disk}$.  We set the 
   FWHM  of the   Gaussian
    at $0.03 r_c$  and take
     $\Sigma(r_c) > \Sigma_{\rm max}(r_c)$.
  Thus we idealize the $t=0$
state of the disk to consist
of two components:
  a narrow  torus
representing material  added from
the secondary,
 and a broad component
($=0.3\Sigma_{\rm max}$)
representing
material left from
the previous outburst.
 In view of the concerns raised by Gammie \& Menou (1998),
    we consider the assumption of a constant and uniform
$\alpha_{\rm cold}$ to describe the quiescent evolution
to be suspect
 and therefore do not
   run complete cycles as in Smak (1998).
  Gammie \& Menou  point out that the large
resistivity of the neutral gas expected in quiescence
   may hamper the efficiency of the magneto-rotational
  instability (MRI), the currently favored
  mechanism for angular momentum transport in 
accretion disks (Balbus \& Hawley 1991, 
                 Hawley \& Balbus 1991, 
  for a review see Balbus \& Hawley 1998).
Unfortunately there is no quantitative way at the present
time to carry out full  time dependent calculations based
on a  physically self-consistent formalism which
  satisfies    Gammie \& Menou
(although Menou 2000 represents a step
in this direction). We must
   assume the existence of an $\alpha_{\rm cold}$ ($=0.02$)
so as to set the critical scalings associated with 
$\Sigma_{\rm max}$.
We examine five values of $\alpha_{\rm hot}$:
0.1, 0.15, 0.2, 0.25, and 0.3.
      We    consider   three  specific systems
and so  adopt values of the central mass $M_{\rm WD}$ and outer disk
radius $r_{\rm disk}$    relevant to each.
   The systemic
      values taken for VW Hyi, U Gem, and SS Cyg
models are $(M_{\rm WD}/\msun,$ $r_{\rm disk}/10^{10}$ cm)
$=(0.63,2.26)$, $(1.26,4.54)$, $(1.19,5.86)$,
    respectively.
  These values come from $M_{\rm WD}$ and $P_{\rm orbital}$
values in Ritter \& Kolb (1998),
where we take $r_{\rm disk} = 0.9 r(L_1)$.  
Therefore we carry out a total of 45 runs:
5 $\alpha_{\rm hot}$ values $\times$ 3 $r_{\rm instig}/r_{\rm disk}$
values $\times$  3 systemic parameter values.
  For each run we compute the absolute visual magnitude $M_V$,
the rate of mass flow within the inward moving heating front,
and the local rate of mass flow at $1.1r_{\rm WD }$ which
we take to be the rate of accretion onto the WD.
  This is used as a tracer of the $EUV$ flux.

\subsection{ Long Term Light Curves  }

Figure 1 shows the results of our 45 trials.
The three panels correspond
to   VW Hyi, U Gem,
and SS Cyg, respectively.
The delay $\Delta t(V-EUV)$
      is measured between the onset
of the rapid  rise in $V$
and the increase in ${\dot M}$
onto the WD.
We define the onset of 
  the rapid increase
 in  $V$ flux by a small
spike in $V$ which accompanies
the end of the stagnation stage.
The horizontal dotted lines
in each panel show the
  observed delay
between the rapid rises in 
     $V$ (from AAVSO and VSS/RASNZ
     observations)
and $EUV$ (from {\it EUVE} observations)
taken from MMB.
  The delays are 0.75 d for VW Hyi,
                 1.25 d for U Gem,
           and   1.5  d for SS Cyg.
 Our computed  
    delay does not scale exactly linearly
with  $r_{\rm instig}/r_{\rm disk}$
   because the effective start of strong heating
which follows stagnation comes after a period of slow
growth of the thermally unstable annulus.
  Thus
 after stagnation ends one is left with a hot annulus
which is fairly broad, so that its inner edge
lies somewhat
              interior to $r_{\rm instig} $.
 This instant of  time marks the onset of the rapid
rise in $V$ and the launching of the heating front.
Figure 2 shows the duration of the stagnation
intervals for the trials in Fig. 1.
  These times are generally longer
than the $V-EUV$ delay time intervals
which immediately follow them.

Figure 3 shows the outburst onset for
our ``best fit'' SS Cyg model $-$
that for which $\alpha_{\rm hot}  = 0.2$ and
 $r_{\rm instig}/r_{\rm disk}= 0.7$.
  (Our preference for $\alpha_{\rm hot}=0.2$
comes   from Smak's [1999] investigation
of the Bailey relation [Bailey 1975]
between outburst decay rate in $V$
      and orbital period for
dwarf nova outbursts.)
One sees a long stagnation stage
caused by the increased specific heat of gas
at $\sim10^4$ K.
     Physically,
the addition or subtraction
of heat goes predominantly
into changing the degree
of partial ionization
of the  gas;
the resultant temperature
change is minimal.
There is a spike in $V$ at $t\simeq 6.1$ d accompanying the
end of stagnation.
The lower panel shows the local rate of mass flow
associated with the inward moving heating front,
on the same scale as the local rate of mass flow 
  within the disk at $1.1 r_{\rm WD}$.
  We calculate this by evaluating ${\dot M}(r)=$
$2\pi r \Sigma v_r$ at the local maximum in $\Sigma(r)$
which defines the heating front.
This flow rate peaks
at $\sim10^{20}$ g s$^{-1}$
just after the onset of the front propagation.
  Although this represents an enormous
       local rate of accretion,
it may have negligible observational 
           consequences
due to the shallow gravitational potential
at large radii.
  It is informative to plot ${\dot M}_{\rm front}$
and ${\dot M}_{\rm WD}$
on the same scale, because one can see the 
direct causal relationship between the two;
the rapid rise in  ${\dot M}_{\rm WD}$
  after the heating front  reaches the inner edge
is a direct consequence of the arrival
   of material carried within the heating spike
at the WD.
  The   amplitude of  ${\dot M}_{\rm WD}$
 is determined by the mass carried in the heating front
spike,  and the subsequent slower rise
  to maximum
  over the next $\sim 0.3$ d
    is determined by the sloshing action
of gas as the outburst density profile $\Sigma(r) \propto r^{-3/4}$
is established.

Figure 4 shows the speed
  of the heating front
  and $h/r$ values within the heating front
for the model from Fig. 3.
 The top panel shows the speed
in units of $\alpha r\Omega (h/r)^n$,
where all quantities are evaluated 
at the local maximum in $\Sigma$ within the
heating front, and $n$ is taken to be either
0, 1, or 2.
  The front speed closely  follows $\alpha r\Omega (h/r)^1$
or  $\alpha c_S$ (local hydrostatic
  equilibrium gives $c_S=\Omega h$),
     confirming previous work by
(Meyer 1984) and Lin et al. (1985).
 Eqn.  (1) of Smak (1998) gives a scaling
 for the ``radial velocity of the accreting matter''
which would correspond to our $n=2$ case.
We can exclude this expression as being the heating
front speed.
  It is representative of the
  general  viscous  flow speed  of gas
within the parts of the disk that are far
from the front,
  however (see Fig. 2 of Cannizzo 1998b).
 Figure 5 shows the front speed 
   and  corresponding values of $h/r$
within the front
 in the SS Cyg models for which 
 $r_{\rm instig}/r_{\rm disk}= 0.7$. 
  The 
      evolution of $h/r$ is virtually identical
for all five curves, so 
that the heating front speed just scales
linearly with $\alpha$.

Figure 6 shows the
 evolution  of $\Sigma(r,t)$ and $T(r,t)$
       for the same model
shown in the previous figures.
Each line is separated by 0.25 d.
Three distinct phases are evident.
(i) 
    The first $\sim6$ d of evolution encompasses a slow
shift from the initial profile.
   The initial torus at $4 \times 10^{10}$ cm
smears and the temperature increases slowly.
It is this
      period of evolution for  which Mineshige (1988)
coined the term ``stagnation''.
Stagnation is a direct consequence of the increase
in the specific heat 
by  a factor $\sim10-50$ for gas at $\sim10^4$ K
(see Fig. 1 of C93b).
(ii) 
   After stagnation has ended,
     a strong spike in $\Sigma(r)$
  develops at the inner edge of the annular heating region,
  and the spike  moves to smaller radii.
   The amplitude of the spike decreases strongly  with decreasing
radius, approaching zero  as $r_{\rm front} \rightarrow r_{\rm WD }$.
This stage lasts  1.5 d and is the direct cause of the $V-EUV$ delay;
this is the time required for the effect of strong heating
(i.e., post-stagnation) to be communicated to the inner edge of the
disk.
(iii)
     Now the entire disk has made the transition from 
neutral to ionized gas, but the flow pattern ${\dot M}(r)$,
although  by now altered significantly from the quiescent profile,
is still out of steady state.
  During  the rest of the evolution the $\Sigma(r)$ distribution
changes in the sense that material at large  radii shifts to smaller
radii, asymptotically approaching $\Sigma(r) \propto r^{-3/4}$
and ${\dot M}(r)\simeq$ constant.

Figure 7 shows the evolution of the ratios
of several terms in the energy equation
 to the heating term during stagnation.
  Specifically, we follow $|C/H|$, $|J_V/H|$, 
                               and $|J_R/H|$,
where $H$ is the heating term $(9/8)\nu\Omega^2\Sigma$,
 $C$ is the cooling term $\sigma T_{\rm eff}^4$,
and $J_V$ and $J_R$ are the viscous and radiative radial flux
terms given earlier.
  One can see the increase with time of the annular extent
  of the thermally unstable ring.
The heating exceeds cooling by a factor $\sim3$ initially
 at the center of the annulus.
  The viscous term $J_V$ is comparable to $C$,
although $|J_V/H|$ is more nearly constant with radius.
By comparing the second to third panels  one sees the 
              insignificance
of the radiative term compared to the viscous  term.
  The relative importance of these two terms has not been
 previously presented in a quantitative way,
  although some workers commented that $J_R$
 is small (e.g. Ludwig \& Meyer 1998). 
Figure 8 shows the evolution of the terms in
the thermal  energy equation $\Delta T =$
$\Delta t ({\rm Term1}-{\rm Term2}-{\rm Term3})$,
where the direct heating and cooling term
       ${\rm Term1}= 2(H-C+J)/(c_p\Sigma)    $,
 the $PdV$ work term   
 ${\rm Term2}= \Re T/(\mu c_p) (1/r) \partial(r v_r)/\partial r $,
      and the advective term
 ${\rm Term3} =  v_r \partial T/\partial r $  (C93b). 
 The $PdV$ term is less than Term1 by about
an order of magnitude, while the advective term 
  is comparable to Term1 in the interior of the stagnation
annulus, and exceeds it slightly near the edges.

\subsection{ Sensitivity to the Initial Mass Distribution }

The calculations of the previous
subsection are based on simplifying
assumptions, adopted for generality,
that the initial mass distribution
in the accretion disk
at the time of triggering
is a sum of two distributions  $-$
a broad one $\Sigma_{\rm broad}$
    representing matter left over
from the previous outburst, and a narrow
one $\Sigma_{\rm narrow }$ 
    comprising matter transferred
from the secondary   star
  which accumulates
  over some annular extent
  between the specific
angular momentum radius
 and the outer disk edge.
  As discussed  previously,
given the uncertainty about the physical
processes which operate during quiescence
and influence the $\Sigma(r)$ evolution,
our parameter study is intended to explore a
broader range of triggering conditions
than might be obtained in previous studies
such as that of Smak (1998) which follow
the evolution of the accretion disk
through several complete cycles under the
assumption of some $\alpha_{\rm cold}$
which is constant in space and time.

We now look at the sensitivity
of our results to the initial 
mass distribution.
The $V-EUV$ delay essentially
represents the time for
the heating front to traverse
the disk from $r_{\rm instig}$ to 
$r_{\rm WD}$, therefore
we need to understand how $v_F$
depends on the background $\Sigma(r)$ profile.
 To explore this we now impose the
additional constraint on our models
 that  $\Sigma_{\rm broad}(r) \le \Sigma_{\rm floor}(r)$,
where in separate runs we take
    $\Sigma_{\rm floor}   $  to be either
      0.01, 0.1, 1, 10, or  100 g cm$^{-2}$. 
 We limit the maximum value of $\Sigma_{\rm broad}(r)$
to $0.7  \Sigma_{\rm max }(r)$.
Figure 9 shows the results for our SS Cyg
model in which $\alpha_{\rm cold} = 0.02$ and
      $\alpha_{\rm hot } = 0.2 $.
The dashed lines indicate the delays from our
previous models.
  One can see that as $\Sigma_{\rm floor}\rightarrow 0$,
the time delays asymptote to a constant value
equal to roughly twice that found earlier
in the results
for which $\Sigma_{\rm broad} = 0.3\Sigma_{\rm max}$.
In our idealization of a background mass distribution
plus a narrow torus, this limiting case corresponds
  to that in which all the disk mass is initially
contained
in the torus.
One might have possibly 
     expected the heating front to die
out as $\Sigma_{\rm floor}\rightarrow 0$;
    we can definitely exclude this.
The strong gradient in viscous stress within the
expanding hot region is the primary agent
which determines the properties of the heating front;
the background medium into which the hot material
expands affects the expansion rate only to second order,
given that the viscous stresses therein are much smaller.
Once $\Sigma_{\rm floor}$ drops below $\sim1$ g cm$^{-2}$,
$(\nu\Sigma)_{\rm floor} << (\nu\Sigma)_{\rm front}$
so that the background profile does  not influence $v_F$.
Figure 10 shows the evolution of $v_F$ for the runs
accompanying Figure 9.
The trials for $\Sigma_{\rm floor} = $ 0.01, 0.1, and 1
  g cm$^{-2}$
are indistinguishable.
For  larger $\Sigma_{\rm floor}$, 
the heating front speed $|v_F|$ increases
because the greater disk mass supports a larger
$\Sigma_{\rm front}$ and therefore larger value
of $(h/r)_{\rm front}$.

Finally, we look at the effect of varying $\alpha_{\rm cold}$
by a factor of 2 around our nominal value of 0.02.
The critical surface density $\Sigma_{\rm max}$
scales as $\alpha_{\rm cold}^{-0.8}$,
therefore larger $\alpha_{\rm cold}$ values
translate into smaller accretion disk masses,
which would lead one to predict smaller amplitude
heating spikes traveling at slower speeds.
Figure 11 indeed shows that
the delays are longer for larger $\alpha_{\rm cold}$,
  however
the dependence is weak:
a factor of 4 increase in $\alpha_{\rm cold}$
only increases the front travel time $\Delta t(V-EUV)$
$=\int dr |v_F|^{-1}$
by $\sim10-20$ \%.
Figure 12 shows the variation in the front speed $v_F$.
The maximum variation in $|v_F|$
occurs at $\sim10^{10}$ cm in these models,
ranging from $\sim3.6$ km s$^{-1}$
to  $\sim2.5$ km s$^{-1}$  as $\alpha_{\rm cold}$ varies
from 0.01 to 0.04.
The variation at larger radii is less,
which accounts for the relatively small range in
$\Delta t(V-EUV)$ as a function of  $\alpha_{\rm cold}$.

\subsection{ Comparison with Menou, Hameury, \& Stehle (1999) }

The most detailed and comprehensive study
 to date which investigates the structure and properties
of transition fronts in accretion disks is that of
Menou, Hameury, \& Stehle (1999=MHS).
These authors
 utilize a fully implicit,
adaptive mesh code (Hameury et al. 1998)
 with  800 radial grid points.
   Their dynamic allocation
of grid points is optimized to place more grid
points in regions with steep   gradients.
This offers a distinct   advantage over a
spatially fixed grid:
their transition fronts generally contain
$\sim100$ grid points and hence are 
well resolved
 even at small radii. 
Insofar as detailed comparisons
can be made between our two studies,
the results appear to be consistent.
For example, in their presentation
of results for ``outside-in'' heating fronts
(see their sects. 3.2 and 3.3)
which contain runs covering a variety of assumptions
for  $\alpha_{\rm cold}$ and   $\alpha_{\rm hot }$,
they find $|v_F|$ values of several km s$^{-1}$
as do we.
In their discussion they comment that  $|v_F|\propto
\alpha_{\rm hot }$, which is  consistent with our
results shown in Figure 5.
The front speeds depicted in MHS are only shown in 
units of km s$^{-1}$,
therefore one cannot quantitatively assess their
departures of  $|v_F|$  from $(\alpha c_S)_{\rm front}$.
Although our Figure 4 shows that
$|v_F| \simeq (\alpha c_S)_{\rm front}$
in  order of magnitude,
our Figs. 5, 10, and 12 which depict
$|v_F|/(\alpha c_S)_{\rm front}$
in more detail reveal a departure
by up to a factor of $\sim3$
in $|v_F|/(\alpha c_S)_{\rm front}$ from unity.
Anticipating future investigations
of  other systematic effects,
we suspect that the scaling
$|v_F| \simeq (\alpha c_S)_{\rm front}$
may hold to no better than a factor $\la10$
 in  regions with the strongest variations.
MHS also mention finding a weak inverse relation
between  $|v_F| $  and  $\alpha_{\rm cold}$  $-$
consistent with our results shown in Figure 12.

Our discussion of the relative importance of the 
different terms in the energy equation
relates to the evolution
of the disk during ``stagnation'' which precedes
the outburst in $V$.
MHS discuss the physical effects and the relative importance
of the different energy terms,
and they show ratios of the terms
to the heating term $H$ at one time during the
propagation of a heating front
after stagnation has ended
(see their Figs. 1 and 2).
Since 
  these times are different,
we cannot directly compare
our results to theirs.
MHS show that,
     aside from a narrow precursor
at the inner edge of the outside-in heating front,
all terms except the cooling term $C$ are negligible.
All plots in MHS showing the ratios of the different
energy terms to $H$ are on a linear scale
which allow  one to see how $H$ is small compared
to the other terms within the narrow precursor on the
leading edge of the heating fronts,
 but prevent
one from assessing the relative
ordering of the energy terms
  within the main body of the front 
  where the terms are of the order
of or smaller than the heating term.

\section {DISCUSSION }

We have examined the delay between the onset of 
visual and $EUV$ fluxes during a dwarf nova
outburst, focusing on three systems for
which complete 
     $V$ and $EUV$ observations exist.
   We run 45 models spanning a range in systemic
parameters, viscosity parameters, and triggering
radii.  When we combine our findings with those
of Smak (1999), who prefers $\alpha_{\rm hot}=0.2$
    based on  the normalization for
the Bailey relation, we infer that the triggering
radii must be close to outer edge of the disk,
at $\sim 0.6-0.7$ of the primary's Roche radius.
  This seems consistent with the time dependent
calculations of Ichikawa \& Osaki (1992) and Smak (1998)
which include a provision for a variable outer disk
radius, in which one sees a contraction of the 
outer disk in quiescence due to  the accumulation
of low angular momentum material from the secondary.

We see a prolonged period of stagnation after the
thermal instability is triggered, during which
time 
    there is only a very gradual increase in the disk temperatures
and a 
      slow spreading of the thermally unstable annulus.
During stagnation one sees    a weak maximum in $\Sigma$
  at the inner edge of the annulus 
which propagates to smaller radii.
The speed of this front is also $\alpha c_S$
(where $\alpha c_S$ is evaluated at the peak of the spike),
just as for the main heating front
  which begins later.
It is slow due to  $\alpha $ and $c_S$ being small.
    Stagnation lasts
       between  $\sim1$ d  and $\sim10$ d
  in our simulations.
Mineshige (1988)
discusses several physical effects which contribute
to stagnation.
   The only effect which appears not to be model
dependent  is that of the increased
specific heat of gas at $\sim10^4$ K due to 
the increased number of degrees of freedom of
partially ionized gas.
 For instance,
Mineshige discusses a contribution to stagnation
due to the steep $T_{\rm eff}(\Sigma)$
dependence near the local maximum in $\Sigma$
in the equilibrium  $T_{\rm eff} - \Sigma $
 relation.
 This results in a near equality between the heating
and  cooling functions
  during the early stages of the thermal instability.
In this work we utilize three simple power law
scalings to describe the equilibrium
$T_{\rm eff} - \Sigma $ relation so as to remain
as model independent as possible (C93b).
Complications in the vertical structure
due to  physics which is not well understood,
such as convection in a strongly sheared medium
(Gu et al. 2000),
        casts doubt on the trustworthiness
of the detailed features  
    of the equilibrium  $T_{\rm eff}(\Sigma)$
relation $-$
    specifically   the double hysteresis
and intermediate stable branch of solutions
 which Mineshige (1988)
  uses in his calculations.

More generally, 
       concerns raised by Gammie \& Menou (1998)
cast doubt on the standard assumption of having a well
defined $\alpha_{\rm cold}$
throughout quiescence. The lack of partial ionization
may impede the operation of the
     MRI  which is thought to give rise to  viscous
dissipation 
        and angular momentum transport
         in accretion disks. 
 Fleming et al. (2000)
 calculate the nonlinear saturation level
 of the MRI as a function of magnetic Reynolds number
$Re_M=c_S H/\eta$,
where $\eta$ is the resistivity,
  for a variety of assumptions concerning
  the initial magnetic field geometry.
  They find that there is no universal constant
  to which $\alpha$  saturates,
but rather one finds a complicated,
model dependent relation between $Re_M$ and $\alpha$
  (i.e. dependence on initial
  magnetic field configuration). 
Furthermore, 
     it  is unclear how strongly dependent
the $\alpha(Re_M)$ values found by Fleming
et al are on their specific computational model.
  There may also be other physical effects that
need to be considered in the 3D MHD  calculations,
such as Hall currents associated
with the weakly ionized quiescent state,
where large resistivity is an issue 
    (Balbus \& Terquem 2001).
 If MHD modelers of the MRI instability can 
reach a consensus on a general $\alpha(Re_M)$ relation,
that will serve as impetus for time dependent
dwarf nova modelers to revisit their calculations.

Mineshige (1988) notes
     that the stagnation stage in his calculations
lasts $\sim1$ d
 and states that
       stagnation
lengthens  the $V-EUV$ delay more
 than in  previous studies
(PVW, CK).
Mineshige suggests that the failure of the previous studies
was tied to their  over-simplified energy equation
$-$ specifically,  their adoption of a constant
  (and small)  $c_P$ value.
This view is not borne out by our calculations.
We find an abrupt increase in $V$ by $\sim5$ mag
lasting a few hours once stagnation has ended.
     Stagnation has no effect on the  $V-EUV$ delay:
  it only influences the evolution prior to the start
of the rapid rise in $V$.
   (Indeed, were we to add the stagnation
intervals $\sim1-10$ d to the heating
  front travel times $\sim1-3$ d,
we would obtain delays considerably 
  longer than observed.
  Mineshige might not have appreciated this
since he only ran one model,
for which the stagnation lasted $\sim1$ d.)
There is a weak  heating front associated with
the spread of the warm region during stagnation,
but the $V$ flux is low during this entire phase.
       After 
       stagnation has ended
       a strong heating front forms
       and $1-3$ d later it reaches
       the WD.
As noted by Smak (1998),
  the failure of earlier works
was caused by their inability 
      to produce outbursts
which were triggered at sufficiently
large radii.
     Although our results seem to contradict
those of Mineshige (1988) by showing that stagnation
has no effect on $\Delta t(V-EUV)$,
  it would probably be short-sighted on our  part to
assert  that stagnation  has no observational
consequences.
Various groups have claimed
  to see  subtle changes in dwarf novae
just prior to   optical outburst (e.g.
Mansperger \& Kaitchuck 1990).
     Our
  simplified 1D treatment
 of the disk equations
and energy flow, 
coupled with our  simplistic methods for computing
the disk fluxes,
    do not currently allow us
to make a firm theoretical prediction
about the observational consequences
of stagnation.

A contraction of the outer disk
during quiescence due to the accumulation of
low angular momentum material, followed by
    an augmentation in the surface density
at large radii is one way to generate type A
outbursts, but it is probably not the only way.
  Even within the context of the formalism
for the treatment of the outer boundary 
condition introduced by Smak (1984)
which was based on earlier work by Papaloizou \&
Pringle (1977),
there are some apparent inconsistencies.
For instance, Ichikawa \& Osaki (1994) present
a comparative study using both   linear and
nonlinear treatments of the tidal torque
 at large radii in the disk.
  They conclude  that the expression utilized by Smak
and later workers (e.g., Ichikawa \& Osaki 1992,
    Hameury et al. 1998)
  in which the tidal torque varies as the
fifth power of radius
does not adequately represent the abrupt nature
of the tidal cut-off at large radii
(see Figs. 3 and 4 of Ichikawa \& Osaki 1994).
In addition, the strength of the tidal torque as
characterized by its linear coefficient
over-predicts the amplitude by about a factor of 20 
  that given by the full nonlinear calculation.
Another consideration as regards type A versus type B
outbursts is the work by  Gammie \& Menou (1998).
They note that due to the low partial   ionization
of the disk gas in quiescence,
the MRI  may be ineffectual,
leading to a state of zero viscosity.
    This view is supported by Menou (2000),
who finds runaway cooling
  during the transition to quiescence
      in a self-consistent
      model using 
 $\alpha(Re_M)$ values from Fleming et al. (2000).
 If there  were no other angular momentum transport 
mechanism operating in quiescence,
      the gas would
pile up at some radius determined primarily by the
specific angular momentum of the gas leaving the
secondary star, and   the triggering
of the thermal instability might be the onset of
some purely hydrodynamical instability
within the accumulating  torus.
  After interior temperatures rise to the point
at which partial  ionization is significant and the
magnetic field entrained in the gas can couple
effectively with the gas,
  the evolution would continue along the
lines of what is presented in this paper.
 Alternatively, Menou (2000) suggests that some
additional process such as spiral shocks may provide
angular momentum transport and accretion during
quiescence.
%

In accord with Meyer (1984) and Lin et al. (1985),
we find
the heating front speed is given
 (to within a factor of $\sim3$)
    by $\alpha_{\rm hot} c_S$,
  where
  the sound speed
    $c_S$ is evaluated within the heating front.
  This gives a delay time $\Delta t(V-EUV)
 =\int dr |v_F(r)|^{-1}  \simeq $
$1.5$ d $(r_s/2.9\times 10^{10} \ {\rm cm})^{1.5}
 (M_1/1.2\msun)^{-1/2}
 (\alpha_{\rm hot}/0.2)^{-1}$,
   where the input values have been scaled to
 those of  SS Cyg,
 and we assume  $h/r=0.01$ independent of radius.
 The ``start'' radius $r_s$ is the point at which the
strong heating front develops following stagnation;
  it is somewhat less than the ``instigation radius''
 $r_{\rm instig}$
because of the slow spreading  of the thermally unstable
annulus during stagnation.
This expression is general and should
be applicable  to other types of interacting 
binaries in which the accretion disk
limit cycle operates $-$
for instance the low-mass X-ray binaries
in which the accreting object is a neutron star
or black hole.
Orosz  et al. (1997)  see a delay of $\sim6$ d
between the rise in $V$ and soft X-ray flux
during the onset of the 1996 April outburst
in GRO J1655-40, a black hole candidate.
Using our scaling for the delay
 and taking $M_1 =6\msun$
gives $r_s=1.3\times 10^{11}$ cm,
which is a considerable fraction of the outer
disk radius $\sim4\times 10^{11}$ cm
expected given the 2.6 d orbital period
$-$
i.e, enough to generate a type A outburst.
This would indicate that  the standard model
can explain the delay for this system
(see however Hameury et al. 1997).

For generality we have adopted the point of view that, while there is good
theoretical and observational evidence for having a well defined
$\alpha_{\rm hot}$ in the outburst state, the viscosity associated 
with quiescence is uncertain. This is unfortunate because the
results presented herein show a significant dependence upon 
$\alpha_{\rm cold}$. If it does turn out to be the case that the viscosity 
in quiescence is non-local (e.g., proceeding via  
spiral shocks) or vanishingly small, 
however,   the whole disk instability  model for dwarf novae would have
to be completely revised. This might then change completely the V-EUV
delay times.  For definiteness we have had to work 
within the framework of the disk instability model since that is the 
only quantitative model which currently exists. For full consistency, 
were one to argue that the current quiescent models are wrong  and then 
advocate an some  alternative (e.g. non-local angular momentum transport),
one would then  have to revise the entire limit cycle model based on the 
new assumptions.  These comments also apply to the  stagnation phase.
Our results  show that the stagnation phase is not detectable in the 
overall light curve.  As mentioned earlier, this may depend on the 
simplified model which  everyone currently uses, but in any event the 
specific values of the stagnation times would be different in a 
drastically different model. Finally,  in this paper we have dealt only
with outside-in outbursts.  Smak (1998) discusses and presents  
computations for inside-out outbursts as well.

\section{ CONCLUSION}

We have presented a time dependent parameter study
of the onset of thermal instability for the accretion
disks in dwarf novae, concentrating on parameters
relevant for three well-studied systems,
VW Hyi, U Gem, and SS Cyg.
     For each system we look at the dependence 
of the delay between the initial rapid increase
in $V$ and $EUV$ fluxes on  triggering radii for
the thermal instability within the disk,
  and the $\alpha$ viscosity parameter for ionized
gas.
 Our basic conclusion is that, for $\alpha_{\rm hot}=0.2$
which was inferred by Smak (1999)
 based on the Bailey relation
  for the decay of dwarf nova outbursts,
we find that triggering must occur at large radii
$\sim0.6-0.7 $ of the WD's Roche radius to produce
delays that are as observed. 
     We find that the speed of the inward
moving heating front which communicates
the effect of the thermal instability is
  $\alpha c_S$ (evaluated within the heating front
spike), in agreement with previous estimates.
We also see a prolonged
period ($\sim1-10$ d)
of slow heating (stagnation)
following the onset of thermal instability
(Mineshige 1988).
In contrast to Mineshige,
we find that stagnation has no effect
on the $V-EUV$ delay since
it precedes the rapid rise in $V$.
  In exploring the sensitivity of our results
to the initial mass distribution, we find
that as the background surface density is made
smaller  (i.e., all the mass is put into the
torus at $r_{\rm instig}$ initially),
the delay $\Delta t(V-EUV)$
 increases by about a factor of two above that
given by calculations in which the background 
surface density $\Sigma_{\rm broad }$
    initially is intermediate
between the critical surface densities
$\Sigma_{\rm min}$  and $\Sigma_{\rm max}$.
  For
   an upper limit background
  surface density
    $\Sigma_{\rm floor} \la 1$ g cm$^{-2}$,
the value of $\Delta t(V-EUV)$ is independent
of $\Sigma_{\rm floor}$.
  If there were a significant amount of
evaporation of the inner disk during quiescence
for instance which depleted the $\Sigma(r)$
values interior to $r_{\rm instig}$
without changing  $r_{\rm instig}$,
that could increase  $\Delta t(V-EUV)$
  by up to a factor $\sim2$ above that found in studies
such as that of Smak (1998).
 A logical next step would be to include some
evaporation at a rate consistent with quiescent
X-ray observations of dwarf novae into
  the  time dependent
calculations of the limit cycle,
  in order to quantify the effect of the depleted
  $\Sigma(r)$ distribution in the inner disk.
It seems unlikely that a self-consistent
 inclusion of this effect would change
 $\Delta t(V-EUV)$ by as much as a factor
of 2,
given that the $\alpha_{\rm hot}$
values of $\sim0.2$
inferred by Smak (1999)
    through independently fitting
both the observed $\Delta t(V-EUV)$
values and rates of decay from outburst
are in agreement.
  Finally, we find a weak inverse dependence of 
  $\Delta t(V-EUV)$ on $\alpha_{\rm cold}$.
     This comes about because larger $\alpha_{\rm cold}$
values lead to smaller accretion disk masses
(via the dependence of $\Sigma_{\rm max}$
on $\alpha_{\rm cold}$),
so that the heating front spike has a smaller
amplitude and lower speed.

We thank Kimberly Engle, Damon Gunther,
 and David Friedlander
              in the 
Laboratory for High Energy Astrophysics
at Goddard Space   Flight Center
for assistance in obtaining large amounts of CPU time
 on various workstations.
We acknowledge useful discussions
  with Ethan Vishniac  and John Hawley.
   We also thank the anonymous referee
whose comments helped focus and improve the paper.
Special thanks go to Chris Mauche whose detailed
discussions  and careful analysis and presentation
of the dwarf nova data
   provided a direct and strong
impetus for carrying out this project.

\vfil\eject
\centerline{ FIGURE CAPTIONS }

Figure 1.  The delay between the initial rapid rise in $V$
    and $EUV$ for our trials which model VW Hyi ({\it top panel}),
  U Gem ({\it middle panel}), and SS Cyg ({\it bottom panel}).
  Each panel contains the results from 15 trials, spanning a
range of 5 values of $\alpha_{\rm hot}$ and 3 values of
  $r_{\rm instig}/r_{\rm disk}$. The values of the central masses,
in solar masses,  are taken to be 0.63, 1.26, and 1.19, respectively,
and the disk outer radii $r_{\rm disk}$, in units of $10^{10}$ cm,
 are 2.26, 4.54, and 5.86, respectively (Ritter \& Kolb 1998). 
The horizontal dotted lines in each panel show the observed
 delay as inferred from observations.

Figure 2. The durations of the stagnation
  period of slow heating  which precedes the rapid rise
  in the $V$ flux, for the simulations shown in Fig. 1.
 The nonlinear dependence of $t_{\rm stagnation}$
on $r_{\rm instig}/r_{\rm disk}$ stems
from the density dependence of the maximum
in the specific heat $c_P(\rho,T)$.

Figure 3. A detailed view of our SS Cyg model, taken from
    the center of the grid of models shown in the bottom panel
of Fig. 1, for which $\alpha_{\rm hot} = 0.2$ and
 $r_{\rm instig}/r_{\rm disk} = 0.7$.
  Shown are the absolute visual magnitude $M_V$ versus time
({\it top panel}) and two   curves giving the local
mass flow rate $2\pi r \Sigma(r) v_r(r)$ $-$
one evaluated  at the peak of the inward moving heating spike,
and one at $1.1r_{\rm WD}$, where $r_{\rm WD} = R(M_{\rm WD})
= 4.2 \times 10^8$ cm for $M_{\rm WD} = 1.19\msun$.

Figure 4. The evolution of the heating front speed
    $v_F$ ({\it top panel})
    and $h/r$  ({\it bottom panel})
     evaluated within the heating front, 
   for the model shown in Fig. 3,
  normalized to  $\alpha r\Omega (h/r)^n$
evaluated at the local maximum in the heating front.
  The best fit to 
  $v_F$ is $\alpha r\Omega (h/r) = \alpha c_S$
  (i.e., that for which the exponent $n=1$).

Figure 5. The evolution of the heating front speed
    $v_F$ in km s$^{-1}$ ({\it top panel}),
  normalized to $(\alpha \Omega h)$
   evaluated at the local maximum in the heating front
  ({\it middle panel}),
    and $h/r$  ({\it bottom panel})
also
     evaluated at the local maximum
     within the heating front,
   for the
  SS Cyg models for which
  $r_{\rm instig}/r_{\rm disk} = 0.7$.
  The five curves shown are for $\alpha_{\rm hot}=$
0.1, 0.15, 0.2, 0.25, and 0.3, with 
      the larger $\alpha_{\rm hot}$ values
corresponding to larger $|v_F|$ values.
     The second panel shows an increasing
deviation in $|v_F|$ from $\alpha c_S$
as $\alpha_{\rm hot}$ increases.
  The value of  $h/r$ within the heating front
is nearly independent of $\alpha_{\rm hot}$.

Figure 6. The evolution of surface density $\Sigma(r,t)$ 
    and midplane temperature $T(r,t)$ over 12 d
for the model shown in Fig. 3.
  The initial profile is  a Gaussian in $\Sigma(r)$
 centered at $r_c \simeq 4\times 10^{10}$ cm.
   Each curve is separated by 0.25 d.
  The 3 stages evident in the evolution are 
(i) a stagnation stage (spanning $\sim6$ d or $\sim25$ curves)
       during which
  the initial  profile broadens by a factor $\sim2-3$, and
$T(r_c)\simeq 10^4$ K,
(ii) a subsequent period of more rapid evolution in which
a strong spike in $\Sigma$ develops and propagates inward
(over $\sim1.5$ d or $5-6$ curves), and 
(iii) a period of readjustment in $\Sigma(r)$ during which
$\Sigma$ at smaller radii is built up as the disk makes a transition
from  being highly out of steady state (i.e., ${\dot M}[r] \neq$ const)
to quasi-steady state (i.e., ${\dot M}[r] \simeq$ const)
 in which $\Sigma$ varies roughly as $r^{-3/4}$.

Figure 7.  The evolution of different terms in the energy
   equation with respect to the heating term for
  the model shown in Fig. 3. 
    The curves represent the state
of the disk at $t$(d)$=1$, 2, 3, 4, and 5,
  which covers the stagnation period. 
  Shown are the logarithms of the absolute values of the
   ratios of (i) the cooling
 term $C$ to the heating term $H$ ({\it upper panel}),
      (ii) the viscous radial flux term $J_V$
       to 
    $H$ ({\it middle panel}),
   and
     (iii) the radiative radial flux term $J_R$ to $H$
          ({\it lower panel}).
  The annular extent of thermal instability
increases with time.

Figure 8. The evolution   of the three terms in the
  energy equation for the same time steps as in Fig. 7.
 Shown are the terms  characterizing the direct, local
heating and cooling  $2(H-C+J)/(c_p\Sigma)$ ({\it upper panel}),
the $PdV$ work term $\Re T/(\mu c_p) (1/r) \partial(r v_r)/\partial r$
  ({\it middle panel}),
and the advective term 
  $v_r \partial T/\partial r$ ({\it lower panel}).
The temperature change in a given time step $\Delta T = \Delta t$
(Term1 $-$ Term2 $-$ Term3), where $\Delta t$
is the time step. The $PdV$ term is less than Term1 by about
an order of magnitude, while the advective term is comparable
to or larger than Term1.

Figure 9. The delay times between the initial rapid rises in $V$
and  $EUV$  fluxes for models in which the broad $\Sigma$ distribution
is not allowed to exceed a value $\Sigma_{\rm floor}$.
 The models adopt the SS Cyg parameters,
 and  the numbers 0.5, 0.7, and 1.0 indicate 
    the value of $r_{\rm instig}/r_{\rm disk }$.
As  $\Sigma_{\rm floor}\rightarrow 0$, the  $ \Delta t(V-EUV)$ value
asymptotes
     to a constant which is about twice that found in our previous
models (indicated by dashed lines).

Figure 10. The heating front speeds and aspect ratios for the models
  shown in Fig. 9. 
  The conventions are the same as in Fig. 5.
The numbers 1, 2, 3, 4, and 5 denote  $\Sigma_{\rm floor}$
values of 0.01, 0.1, 1, 10, and 100 g cm$^{-2}$, respectively.

Figure 11.  The delay times between the initial rapid rises in $V$
and  $EUV$  fluxes as a function of $\alpha_{\rm cold}$.
   The models adopt the SS Cyg parameters,
 and  the numbers 0.5, 0.7, and 1.0 indicate 
    the value of $r_{\rm instig}/r_{\rm disk }$.

Figure 12. The heating front speeds and aspect ratios for the models
  shown in Fig. 11.
    The conventions are the same as in Figs. 5 and 10.
 The five   curves in each panel represent $\alpha_{\rm cold} =$
0.01, 0.015, 0.02, 0.03, and 0.04.

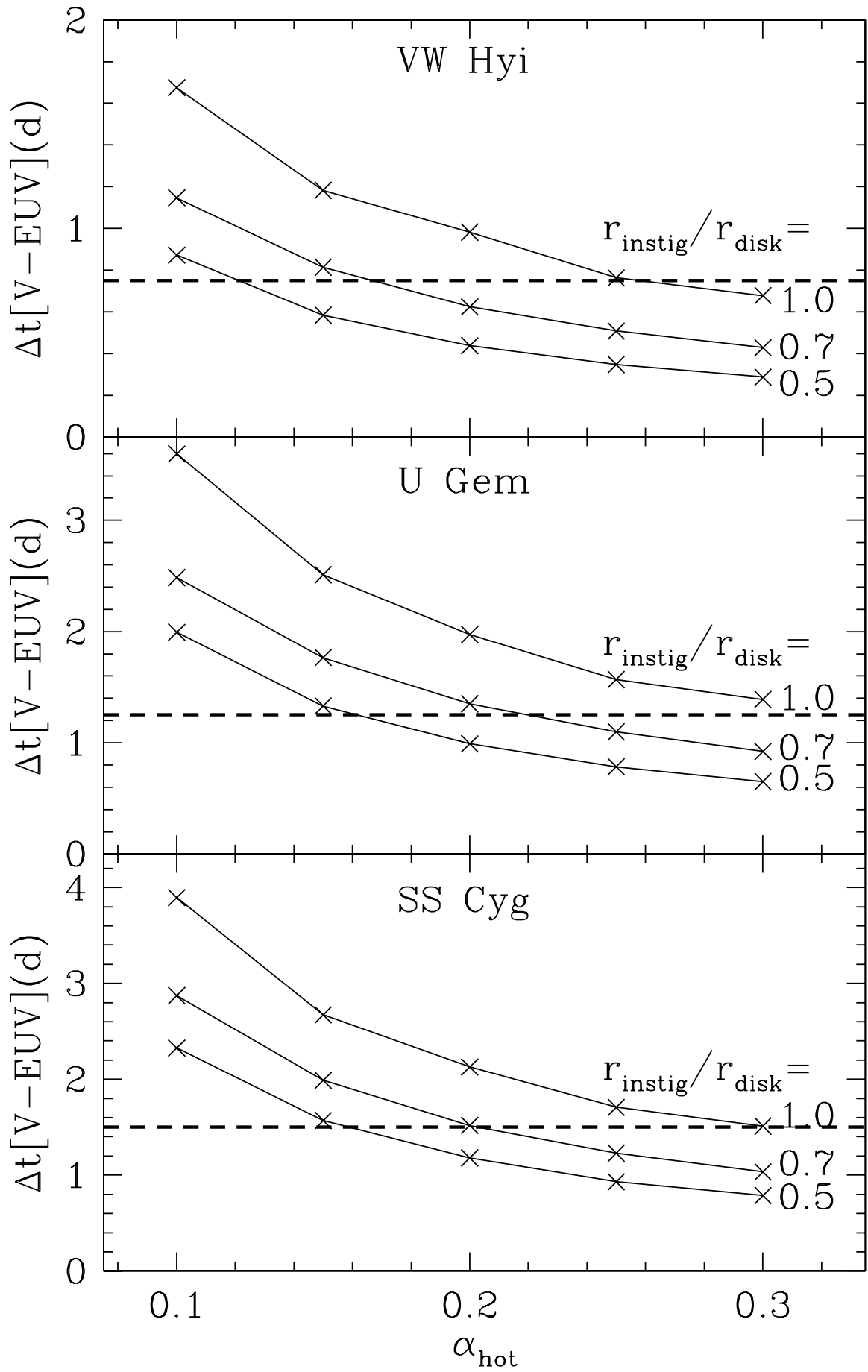
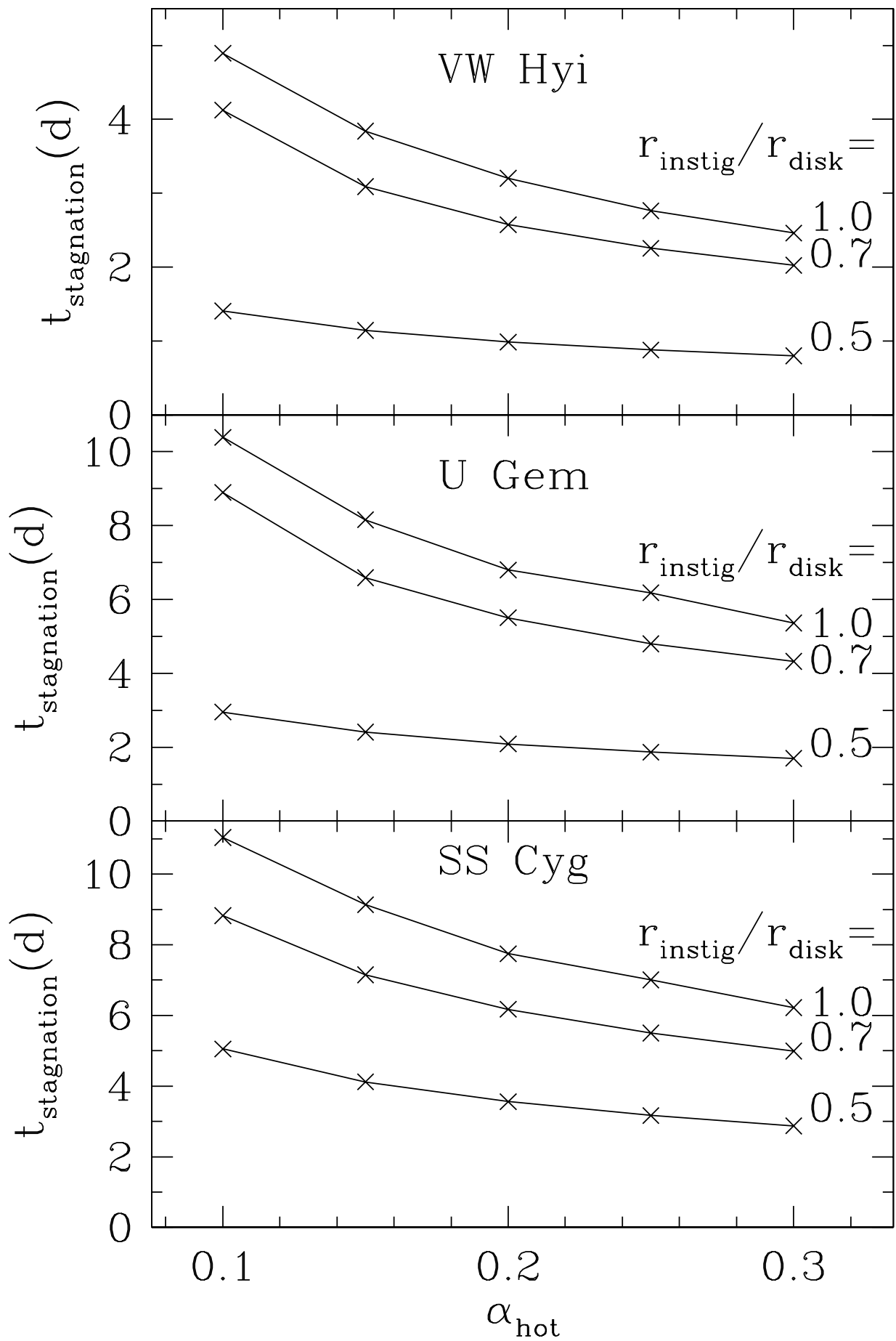
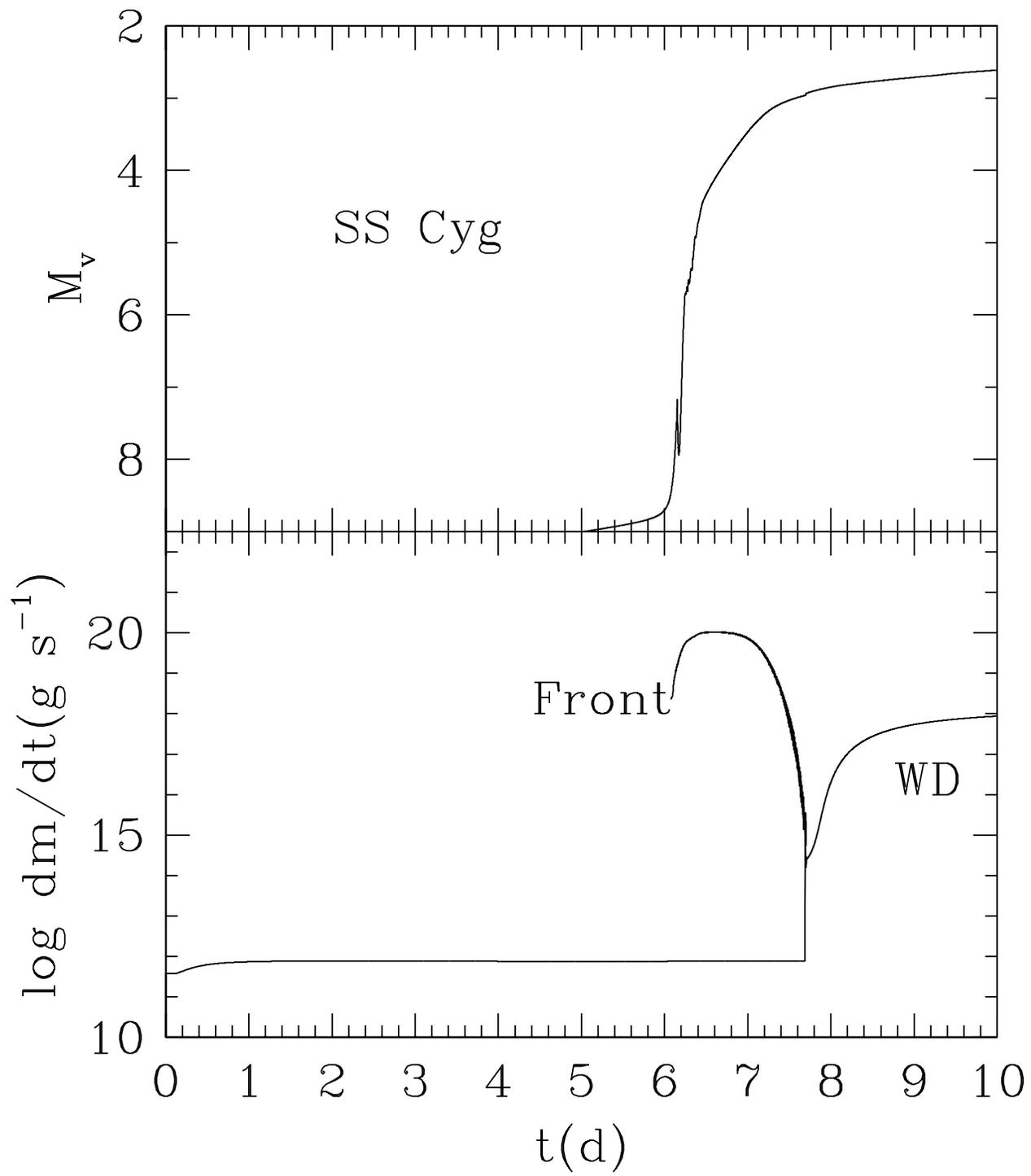
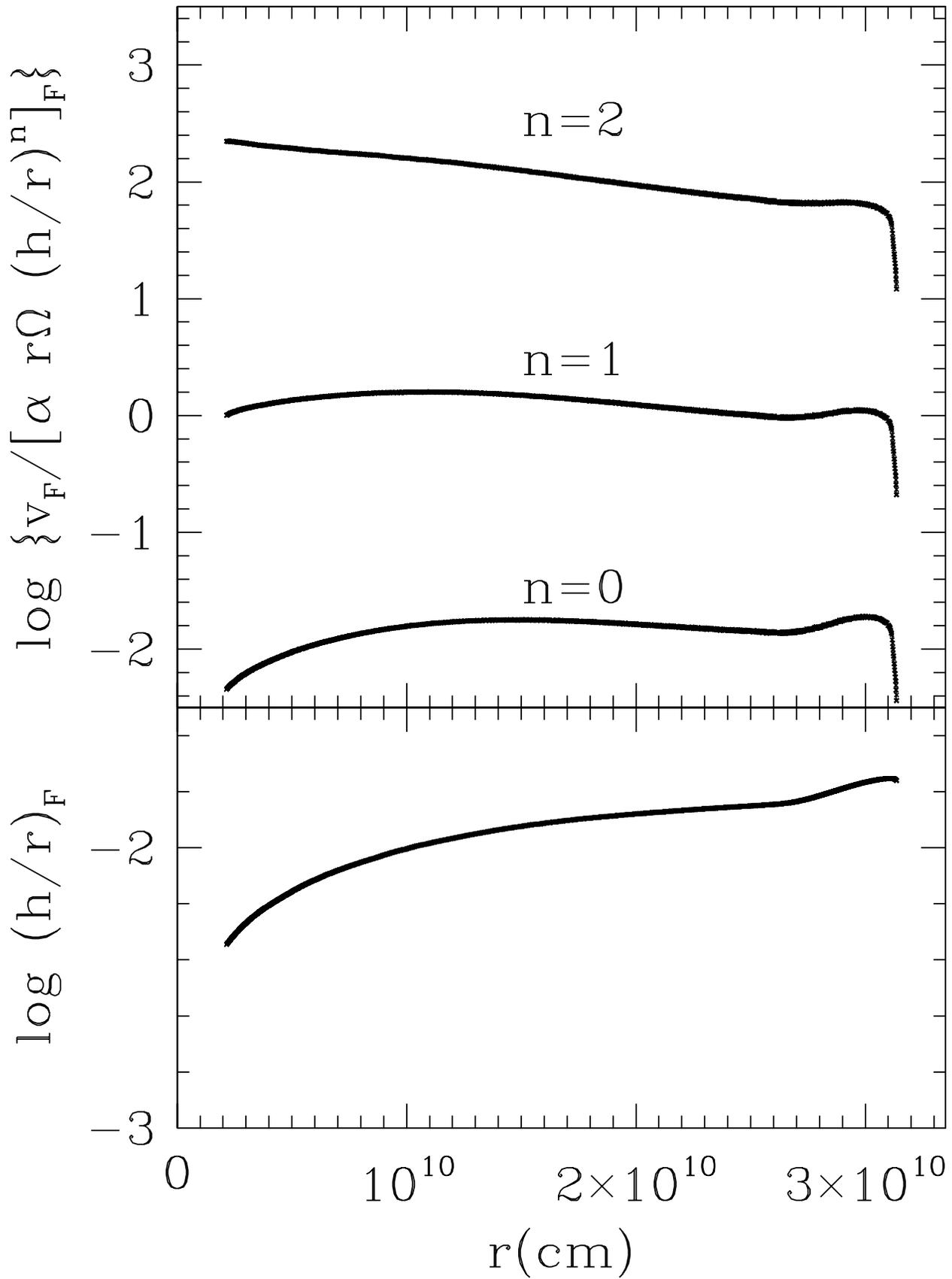
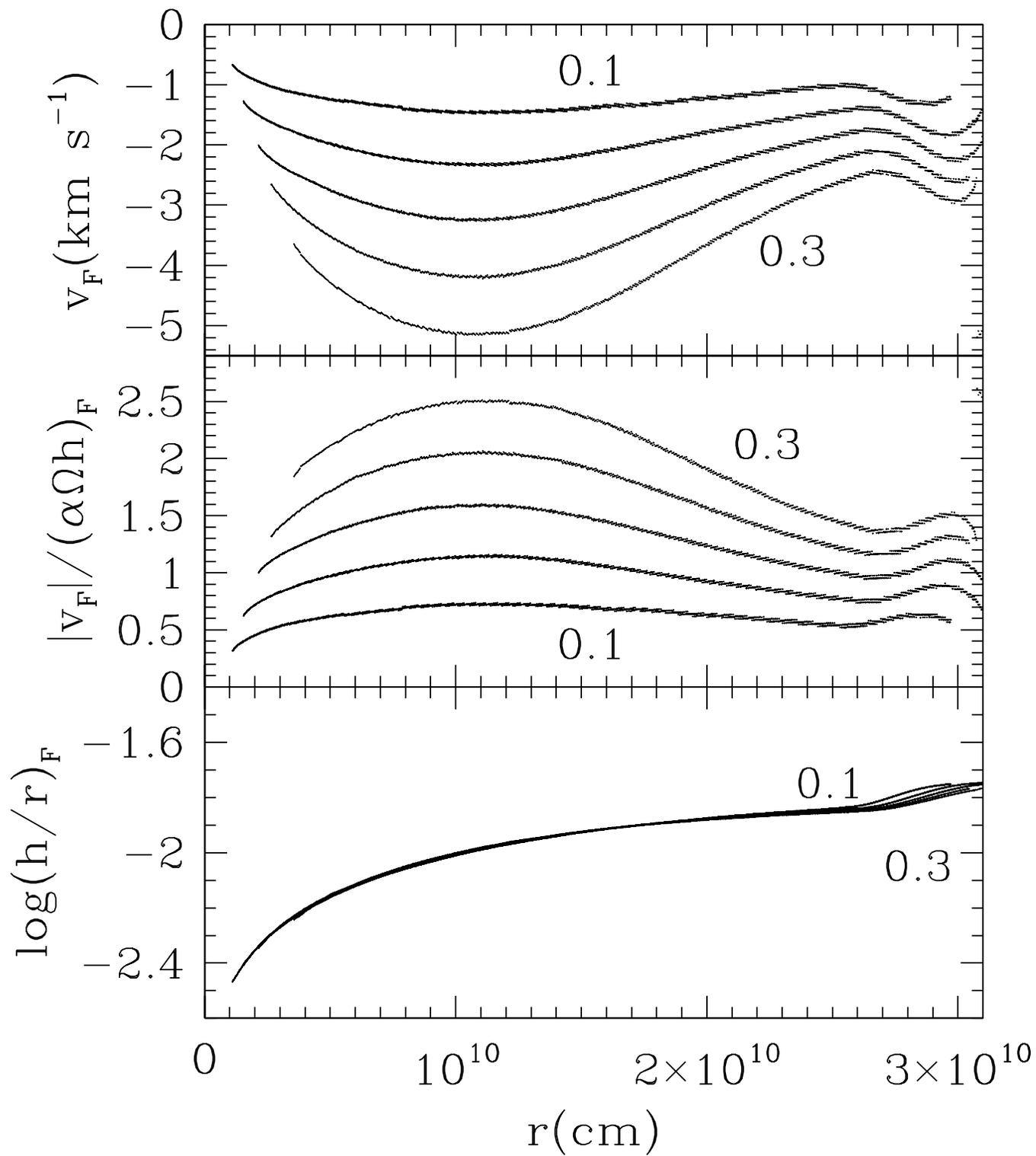
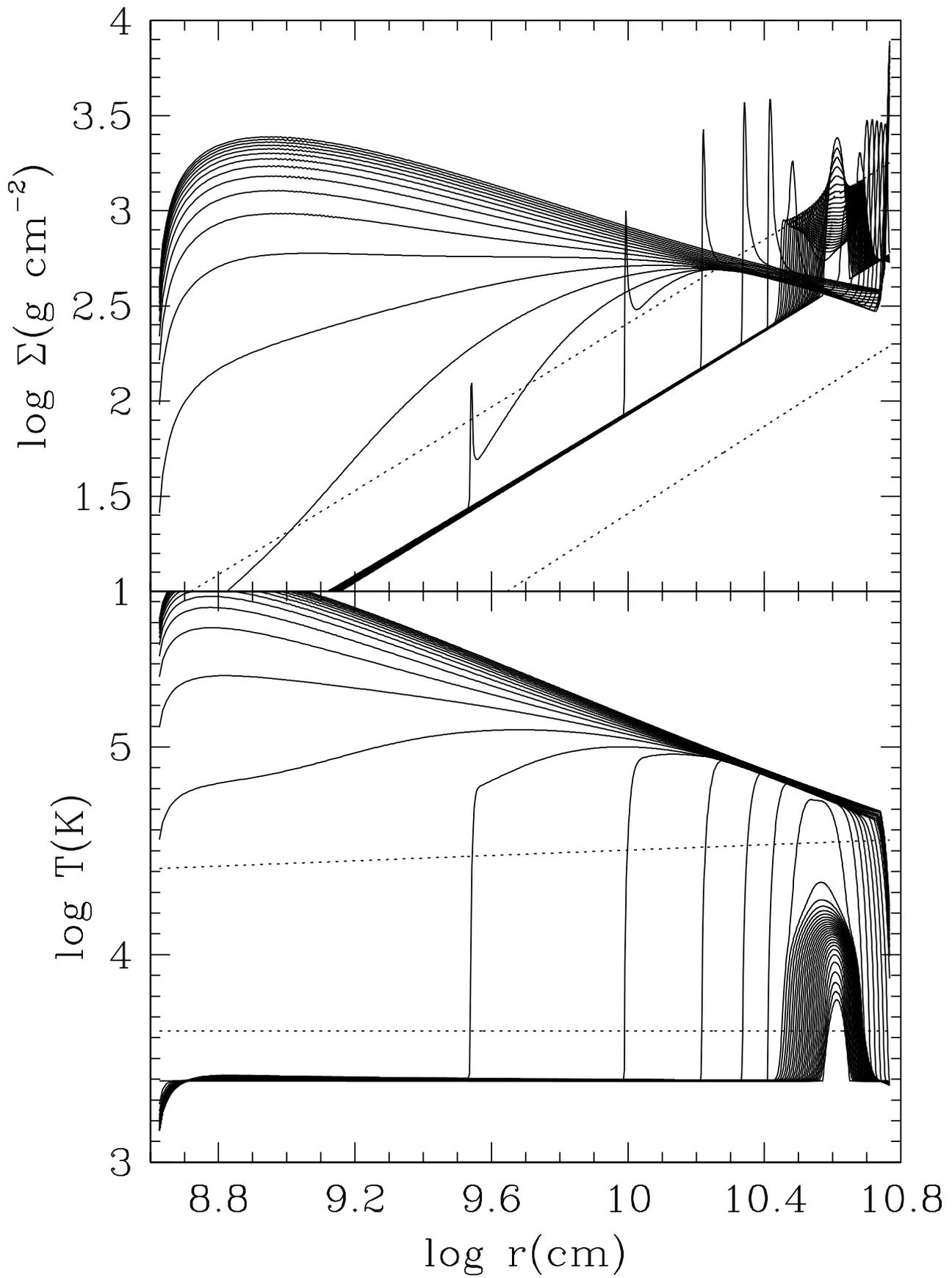
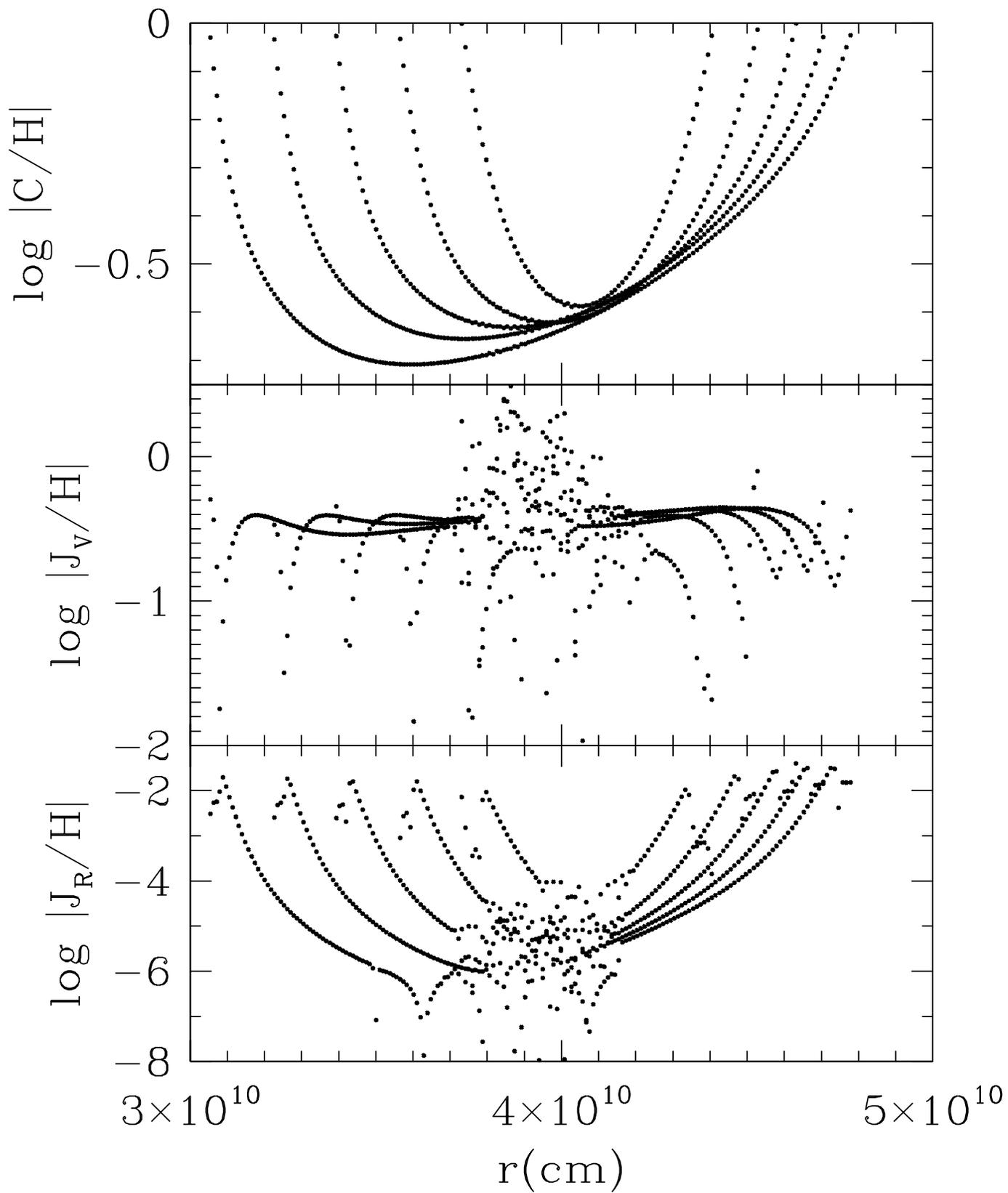
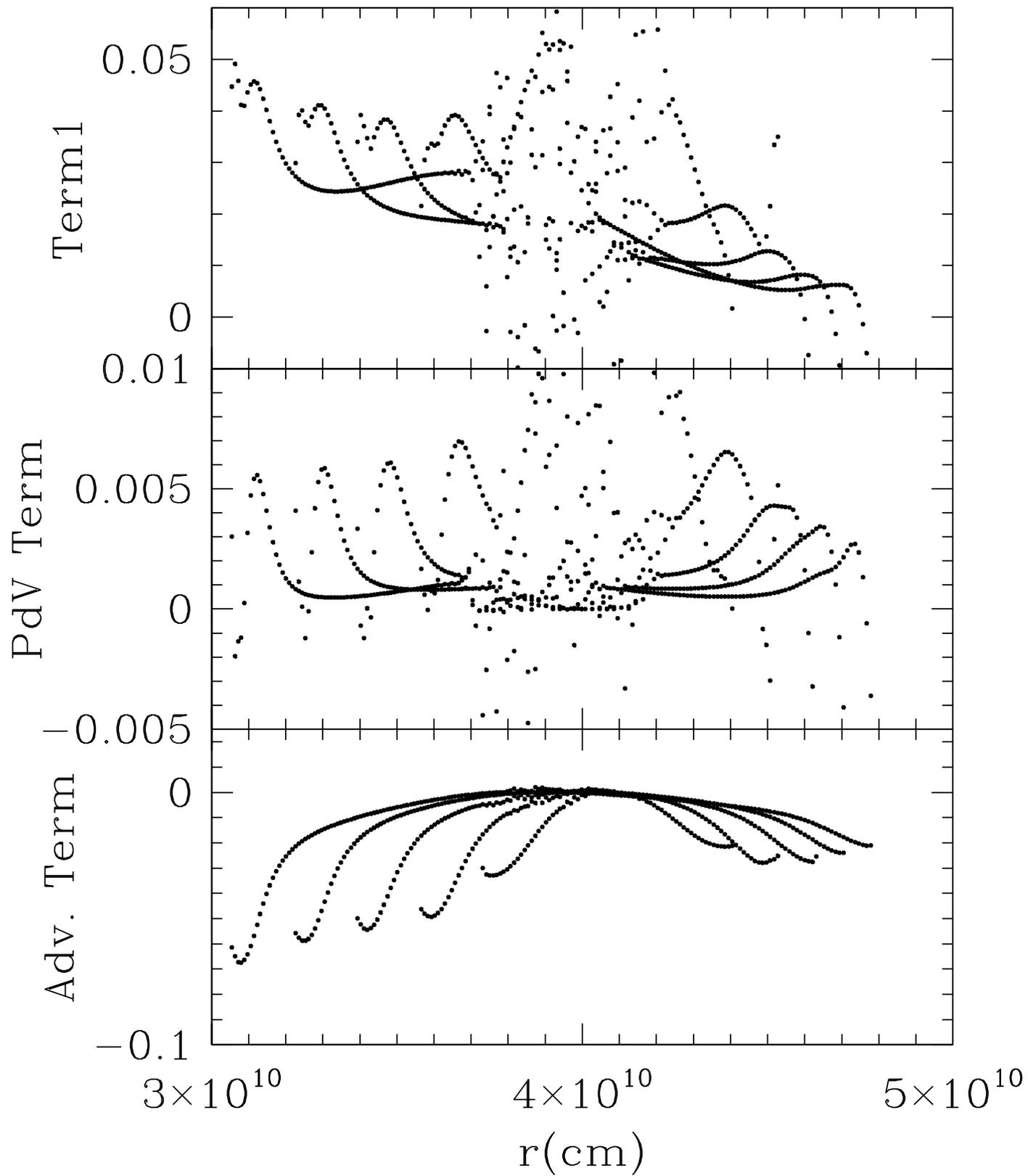
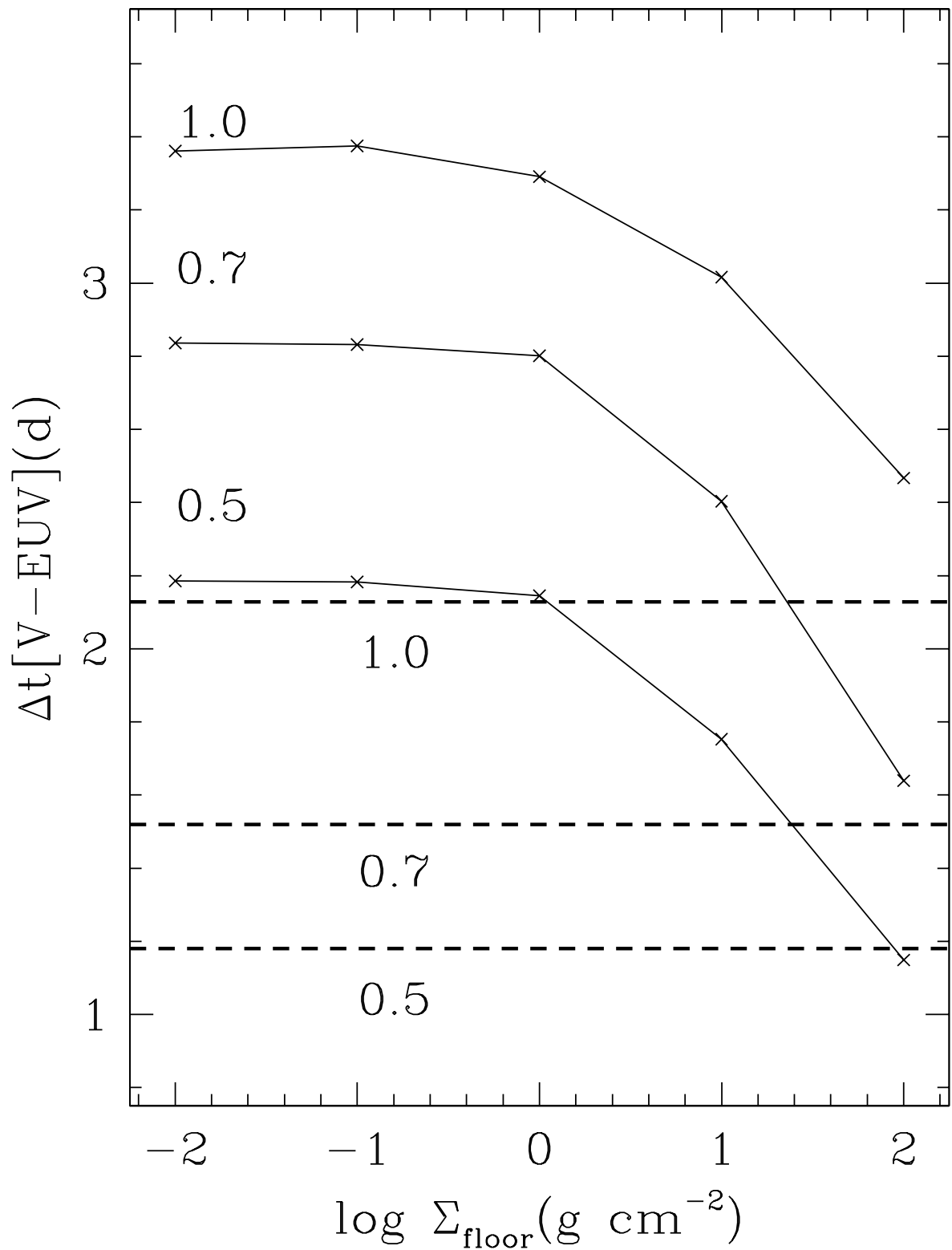
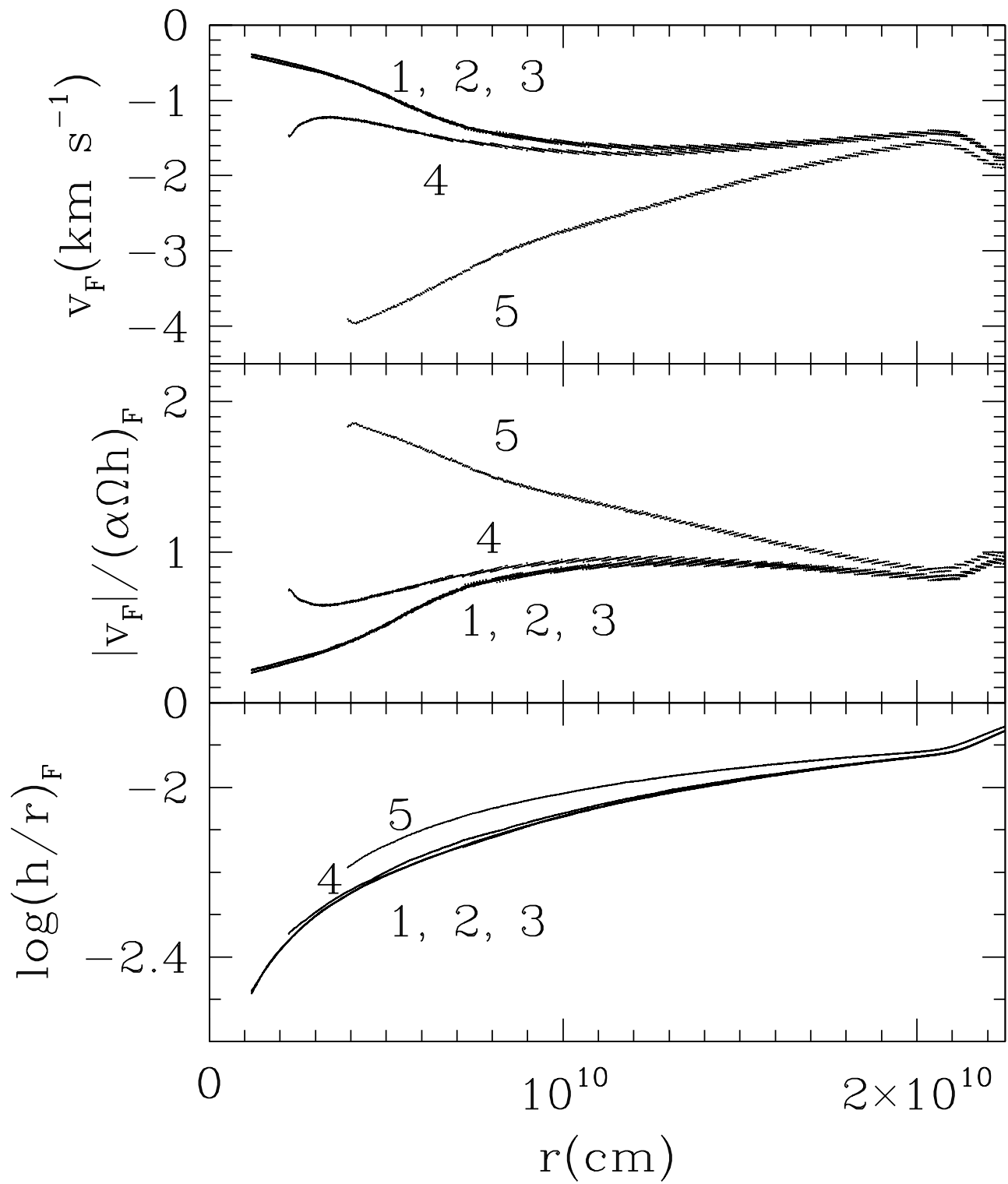
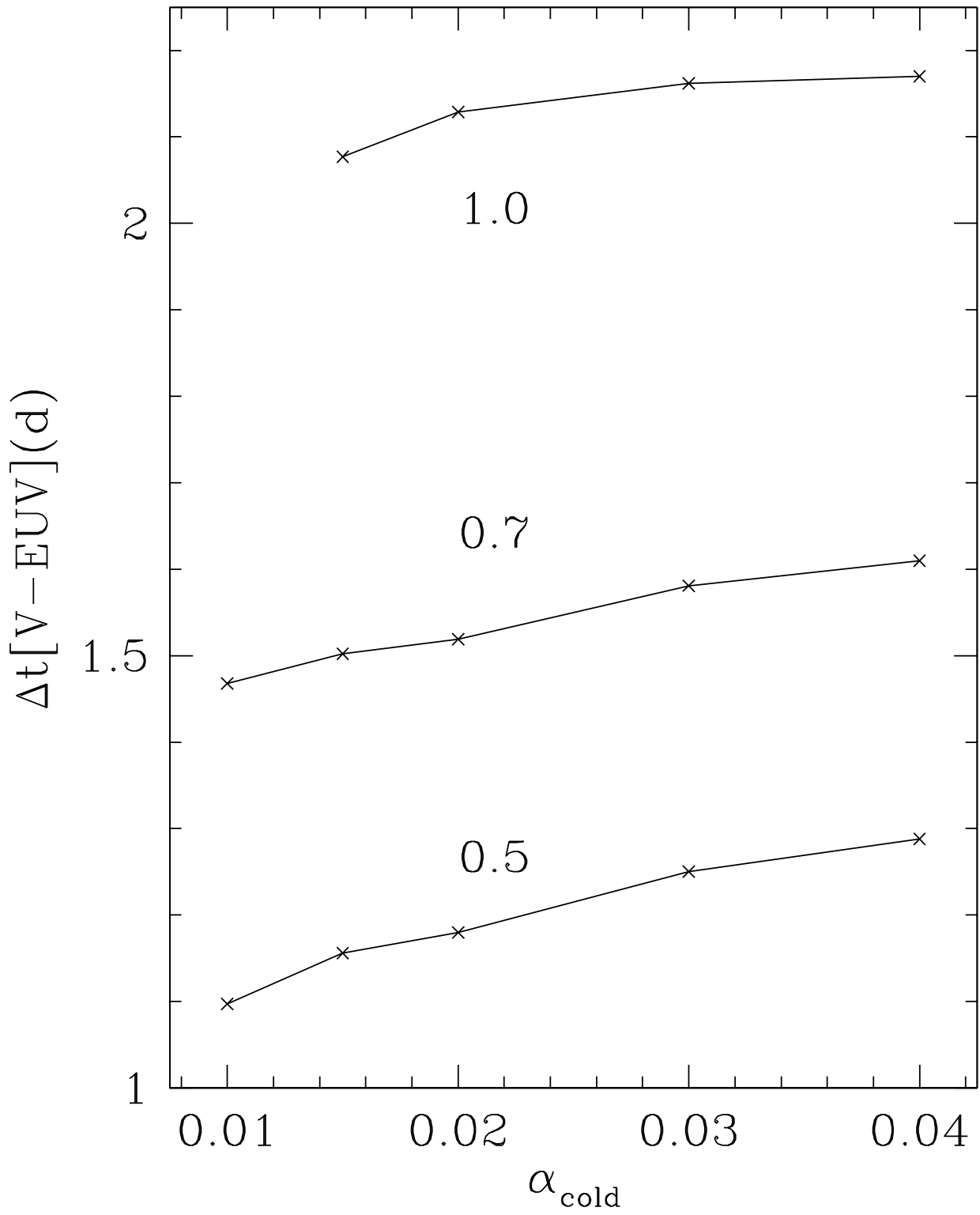
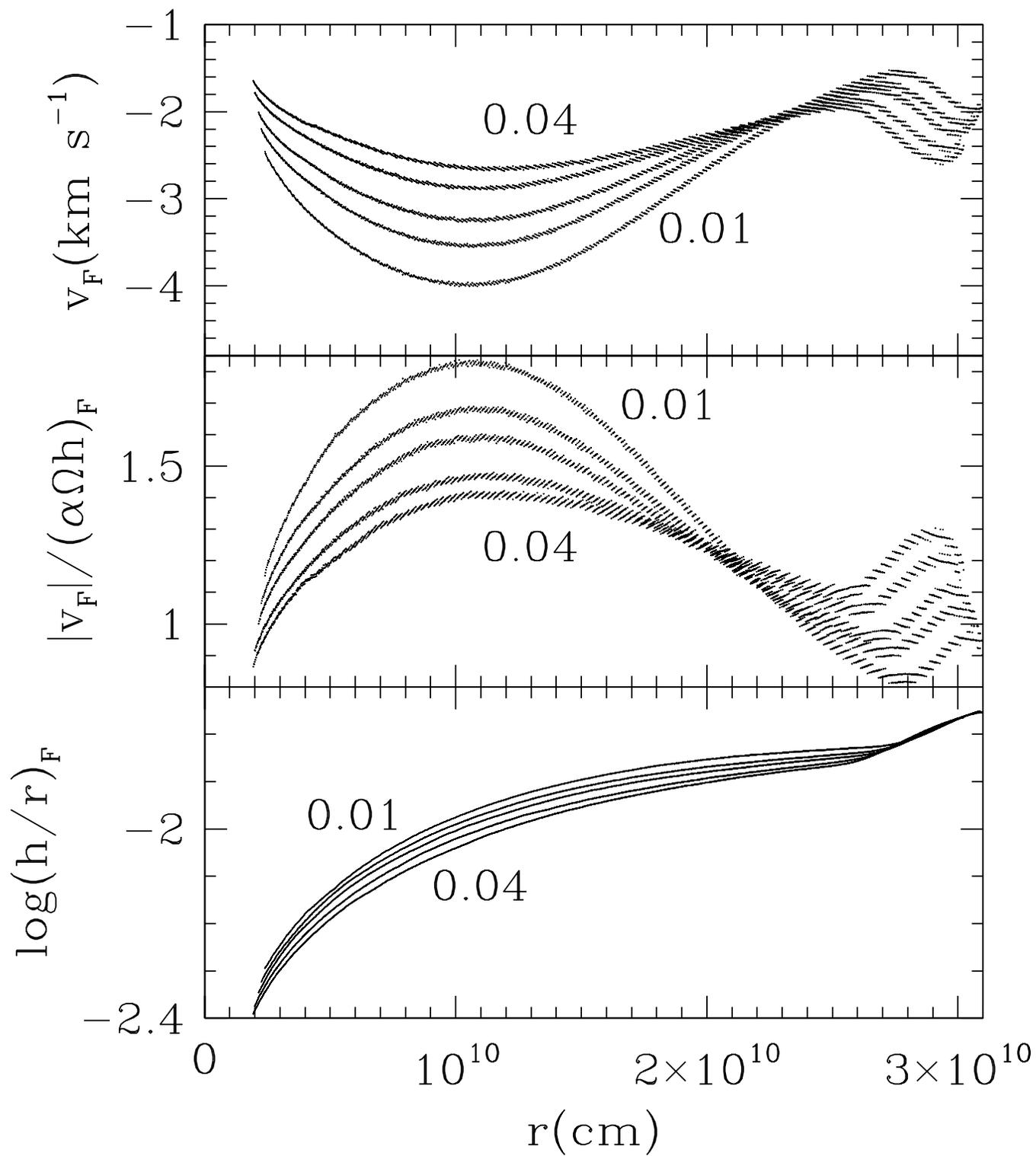
\end{document}